\documentclass[a4paper,11pt]{article}

\usepackage{amssymb}  % MikTeX
\usepackage{graphicx}
\usepackage[usenames]{color}
\usepackage[utf8]{inputenc}
\usepackage{hyperref}
\usepackage{amsmath}
\usepackage{amssymb}
\usepackage{authblk}
\usepackage{lipsum}

\date{}

 \newtheorem{theorem}{Theorem}
\newtheorem{lemma}{Lemma}
\newtheorem{proposition}{Proposition}
\newtheorem{corollary}{Corollary}

 \newcommand{\Keywords}[1]{\par\noindent
{\small{\it Keywords\/}: #1}}

\newcommand{\supp}{{\rm supp}}
\newcommand{\wt}{{\rm wt}}

\begin{document}

	\title{On the Fourier Entropy–Influence Conjecture for Boolean Plateaued  Functions}
	\date{}%leave it blank

\author{Vladimir N. Potapov }

%\affil[*]{\small Faculty of Engineering and Natural Sciences, Sabanc{\i} University, 34956, Istanbul,T\"{u}rkiye,\hskip20mm e-mail:  %quasigroup349@gmail.com}
	
	\maketitle
	\begin{abstract}
	We prove the following inequality for Boolean functions: $$2\sum\limits_{x\in
\mathbb{F}_2^n}f(x)\wt(x)\geq \wt(f)(n-\deg_{alg}(f)).$$
Using this inequality, we establish the Fourier Entropy--Influence (FEI) conjecture for Boolean plateaued functions. In particular, we show that the sharp FEI constant for the class of plateaued functions is $4$. We also prove the FEI conjecture for partially bent functions and show that the corresponding sharp constant is $2$. Finally, we derive several estimates for the $p$-biased distribution on the Boolean hypercube.

\Keywords Fourier entropy, total influence, average sensitivity, plateaued function, algebraic degree, Reed--Muller code, $p$-biased distribution.

	\end{abstract}

	\thispagestyle{empty}

\section{Introduction}

Let $f:\mathbb{F}_2^n\rightarrow \mathbb{F}_2$ be a Boolean
function. Throughout the paper, we use the {\it Walsh--Hadamard (Fourier) transform} of
$f$, normalized by the factor  $2^{-n}$. It is defined by
$$W_f(z)=\frac{1}{2^n}\sum\limits_{x\in
\mathbb{F}_2^n}(-1)^{f(x)\oplus(x,z)},$$ where
$(x,z)=x_1z_1\oplus\cdots\oplus x_nz_n$. Thus, $W_f:\mathbb{F}_2^n\rightarrow
\mathbb{R}$.
 By Parseval's identity,
$\sum\limits_{z\in \mathbb{F}_2^n}W_f^2(z)=1$. 
Therefore,  the values  $W_f^2(z)$ define a 
probability distribution on  $\mathbb{F}_2^n$. The {\it Fourier entropy} of $f$ is defined by $H(f)=\sum\limits_{z\in
\mathbb{F}_2^n}W_f^2(z)\log_2\frac{1}{W_f^2(z)}$. The quantity
$I(f)=\sum\limits_{z\in \mathbb{F}_2^n}W_f^2(z)\wt(z)$ is called the  {\it total influence} or
{\it average sensitivity} of $f$,  where $\wt(z)$ denotes the Hamming weight
of $z$.

The   Fourier Entropy–Influence (FEI) conjecture was introduced by Friedgut and Kalai
 in 1996 \cite{FG}. The conjecture asserts the existence of a universal constant $C$ such that  
 $H(f)\leq C \cdot I(f)$ for every
Boolean function $f$.
 Restricted versions of the FEI conjecture have been proved for several important classes of Boolean functions (see \cite{Li-Tang}). In particular, the FEI conjecture is true for
monomial Boolean functions with $C = 4$,
which is the best possible (see \cite{Stanica}).  Recently, Li and Tang \cite{Li-Tang} constructed $s$-plateaued functions whose FEI constants are strictly smaller than $4$, but arbitrarily close to $4$.

In the present paper, we prove the following inequality
\begin{equation}\label{feie1}
2\sum\limits_{x\in \mathbb{F}_2^n}f(x)\wt(x)\geq \wt(f)(n-\deg_{alg}(f)) ,
\end{equation}
 which relates the average Hamming weight of the support of a Boolean function to its algebraic degree. 
 Here $\wt(x)$ is the Hamming weight of $x\in\mathbb{F}_2^n$ and  $\wt(f)=\sum\limits_{x\in \mathbb{F}_2^n}f(x)$,  i.e.,
  $\wt(f)$ is the total number of values $1$ of a Boolean function $f$.

Using this inequality, we prove that the FEI constant for plateaued functions is at most $4$  (Theorem \ref{feith3}). Combined with the recent construction of plateaued functions whose Fourier min-entropy/influence ratios approach $4$ \cite{Li-Tang}, this shows that the sharp FEI constant for the class of plateaued functions is exactly $4$.
 Furthermore, we prove the FEI conjecture for partially bent functions and show that the sharp FEI constant in this case is equal to $2$ (Theorem \ref{feith2}). Additional relations between total influence and algebraic degree can be found in \cite{Pot}.
 
Local probabilistic testing is a powerful instrument to determine an algebraic degree of Boolean functions \cite{Sala}. 
Codewords of  a Reed--Muller code of order $d$ correspond to  Boolean functions of degree at most $d$. A code is locally testable if there is a way to indicate with high probability that a vector is far enough from any codeword by checking only a small number of the vector's bits. In \cite{Alon} authors proved  that the Reed--Muller codes of constant order are locally testable. 
   In \cite{Dinur} the authors proposed to use  the $p$-biased distribution $\mu(p)$ instead 
   the standard uniform distribution $\mu(1/2)$ distribution on the Boolean hypercube. In Section 5 we consider how useful new inequality for an estimating of probabilities under $\mu(p)$.

\section{FEI Conjecture for Plateaued Functions}

A Boolean function is called $s$-{\it plateaued} if
$|W_f(z)|=2^{\frac{s-n}{2}}$ (equivalently $W_f^2(z)=2^{s-n}$) or
$0$ for every $z\in \mathbb{F}_2^n$. In the case $s=0$,  an $s$-plateaued function is called  {\it bent}.
An $s$-plateaued function $f$ is called  {\it partially bent} function  if 
 $\supp(W_f)$ is an affine subspace.
 The following statements are immediate
 
\begin{proposition}\label{feip1} 
$H(f)=n-s$ for each $s$-plateaued function $f$.
\end{proposition}
Proof. Since every nonzero Walsh coefficient satisfies  $W_f^2(z)=2^{s-n}$, and the support size equals 
$2^{n-s}$, we obtain $H(f)=2^{s-n}\cdot 2^{n-s}\cdot (n-s)$. $\square$

\begin{corollary} $H(f)=2I(f)$ holds for every bent function $f$.
\end{corollary}

It is well known that all directed derivatives of any bent function are balanced. In \cite{Stanica}, the FEI conjecture was proven with constant 
$2$ under weaker conditions: it is sufficient that an 
$n$-variable function possesses 
$n$ linearly independent directions whose derivatives are balanced. 

\begin{proposition}\cite[equality (2.41)]{Carlet}\label{feip2}  For any face
$\Gamma\subset \mathbb{F}_2^n$, $0\in \Gamma$, and a Boolean function
$f$ it holds $$\sum\limits_{z\in
\Gamma}W_f(z)=2^{\dim(\Gamma)-n}\sum\limits_{x\in
\Gamma^\perp}(-1)^{f(x)}.$$
\end{proposition}

 Every Boolean function $f: \mathbb{F}_2^n
\rightarrow \mathbb{F}_2$ can be represented in  {\it algebraic
normal form} (ANF)
\begin{equation}\label{eqZhegal}
f(x_1,\dots,x_n)=\bigoplus\limits_{y\in
    \mathbb{F}_2^n}M_f(y)x^{y}=\bigoplus\limits_{y\in
    \mathbb{F}_2^n}M_f(y)x_1^{y_1}\cdots x_n^{y_n},
\end{equation}
     where $x^0=1, x^1=x$,
and the function $M_f:\mathbb{F}_2^n\rightarrow \mathbb{F}_2$ is
called the {\it M\"obius transform} of $f$.
$\deg_{alg}(f)$ is the maximum degree of the polynomial (\ref{eqZhegal}).
 It is well known (see
\cite[Section 2.2.1]{Carlet}) that for every function the ANF is
unique.

\begin{proposition}\label{feip22} 
Let $f: \mathbb{F}_2^n
\rightarrow \mathbb{F}_2$. $|\supp(f)|$ is odd if and only if $\deg_{alg}(f)=n$.
\end{proposition}
This proposition follows from the definition of ANF immediately.

It is well known that 
\begin{equation}\label{eqMob}
M_f(y)=\bigoplus\limits_{x\in
\Gamma_y}f(x),
\end{equation}
 where the face   $\Gamma_y$, $\dim(\Gamma_y)=\wt(y)$, contained  $0$ and $y$. 
 Let $\geq$ be a natural partial order on $\mathbb{F}_2^n$, i. e., for $x,y\in \mathbb{F}_2^n$ inequality $x\leq y$ implies that $y_1=1$ if $x_i=1$.
 Note that $\Gamma_y$ is the set of vectors $x$ such that $x$ is less than $y$ in this  order. 

 The  maximal degree of the monomial in ANF of
$f$ is called the {\it algebraic degree} of $f$, i.\,e., ${
\deg}_{alg}(f)=\max\limits_{M_f(y)=1} \wt(y)$.

Let $S_f:\mathbb{F}_2^n\rightarrow \mathbb{F}_2$ be an indicator
function of $\supp(W_f)$.

\begin{proposition} If $f$ is an $s$-plateaued function then
$\deg_{alg}(S_f)\leq \frac{n+s}{2}$.
\end{proposition}

Proof. Let $\Gamma\subset \mathbb{F}_2^n$ be a face containing the  zero vector.
By Proposition \ref{feip2} we see that if $\dim(\Gamma)-n> \frac{s-n}{2}$ then
$\Gamma$ contained zero or even number of   nonzero  coefficient $W_f$. Then we obtain
$\deg_{alg}(S_f)\leq \frac{n+s}{2}$ by using (\ref{eqMob}). $\square$

\begin{theorem}\label{feith2} If $f$ is a partially bent function then $H(f)\leq
2I(f)$.
\end{theorem}

Proof. 
Let $f$ be a $s$-plateaued  and partially bent function. Then $S_f$ is an $(n-s)$-affine subspace by the definition. Let $h$ be an indicator function of a $k$-dimensional affine subspace then $\deg_{alg}(h)=n-k$.
By (\ref{feie1}) $2\sum\limits_{x\in
\mathbb{F}_2^n}h(x)\wt(x)\geq \wt(h)(n-(n-k))$ Then 
$$I(f)=\sum\limits_{z\in \mathbb{F}_2^n}W_f^2(z)\wt(z)=2^{s-n}\sum\limits_{z\in  \mathbb{F}_2^n}S_f(z)\wt(z)\geq 2^{s-n}\wt(S_f)\frac{n-s}{2}=\frac{n-s}{2}.$$ Since $H(f)=n-s$ by Proposition \ref{feip1}, we complete the
proof. $\square$

\begin{theorem}\label{feith3}  $H(f)\leq
4I(f)$ for every plateaued function $f$.
\end{theorem}

Proof. $W_f^2(z)=2^{s-n}$ or $W_f^2(z)=0$ for any $z\in
\mathbb{F}_2^n$. By Parseval's identity $\wt(S_f)=2^{n-s}$. By (\ref{feie1})
 we have that
$$I(f)=\sum\limits_{z\in \mathbb{F}_2^n}W_f^2(z)\wt(z)=
2^{s-n}\sum\limits_{z\in \mathbb{F}_2^n}S_f(z)\wt(z)\geq
\frac{n-\deg_{alg}(S_f)}{2}.$$ By Proposition \ref{feip2} we have $I(f)\geq
\frac{n-s}{4}$. Since $H(f)=n-s$ by Proposition \ref{feip1}, we complete the
proof. $\square$

Note that in \cite{Li-Tang}, the authors proved that  $\sup \frac{H(f)}{I(f)}\geq 4$, where the supremum is taken over all plateaued functions. Consequently, the FEI constanf $4$  is tight for plateaued functions.

\section{Proof of the Main Inequality}

Let $S\subseteq \mathbb{F}_2^n$ and $u,v\in \mathbb{F}_2^n$. Define 
$T_S(a,b)=\{x\in S, x\geq a, x\geq b\}$.

\begin{lemma}\label{lemIA1}
For every $S\subset \mathbb{F}_2^n$ there exists a permutation $\pi: S\rightarrow S$ such that  for every $y\in S$ the cardinality $|T_S(y,\pi(y))|$ is odd.
\end{lemma}

Proof.  
Suppose that elements of $S$ are enumerated by integers such that if $x< y$ then the number of $x$ is less than the number of $y$.
Define a matrix $M$ of size $|S|\times |S|$ by the equation
$$ M_{xy}=\left\{
\begin{array}{ll}
1 & \mbox{if}\  x\leq y,\\
0 & \mbox{otherwise}.
\end{array}
\right.$$
We consider $M$ as a matrix over $\mathbb{F}_2$.
By  definition, $M$ is an upper triangle matrix with units on the main diagonal. Therefore, $\det(M)=1$ and $\det(MM^{\mathsf T})=1$. So, there exists a permutation $\pi$ such that $ MM^{\mathsf T}(y,\pi(y))=1$ for every $y\in S$. 

By  definition $$(MM^{\mathsf T})_{uv}=\langle {\bf 1}_{S\cap\{u\leq y\}},  {\bf 1}_{S\cap\{v\leq y\}}\rangle= 
|T_S(u,v)| \bmod 2. $$  Thus, the cardinality $|T_S(y,\pi(y))|$ is odd  for every $y\in S$.  $\square$

\begin{lemma}\label{lemIA2}
 Let $f:\mathbb{F}_2^n\rightarrow \mathbb{F}_2$ be a Boolean function and let $|T_{\supp(f)}(u,v)|$ be odd. Then
$\wt(u)+\wt(v)\geq n
-\deg_{alg}(f).$
\end{lemma}

Proof. 
Consider the  indicator  function $h$ of $T_{\supp(f)}(u,v)$. By definition $h(x)=f(x)x^ux^v$ because of $x^z$ is an indicator function of the set $\{x\in \mathbb{F}_2^n : x\geq z\}$. Therefore, $\deg_{alg}(h)\leq 
\deg_{alg}(f)+\wt(u)+\wt(v)$.  By Proposition \ref{feip22} $\deg_{alg}(h)=n$. Then $\wt(u)+\wt(v)\geq n-\deg_{alg}(f)$. $\square$

\begin{theorem}\label{ThAI3}
 Let $f:\mathbb{F}_2^n\rightarrow \mathbb{F}_2$ be a Boolean function then 
 $$2\sum\limits_{x\in
\mathbb{F}_2^n}f(x)\wt(x)\geq \wt(f)(n-\deg_{alg}(f)).$$
\end{theorem}

Proof.  Let $S=\supp(f)$. By Lemmas \ref{lemIA1} and \ref{lemIA2} there exists a permutation $\pi$ of $S$ such that  $$\wt(y)+\wt(\pi(y))\geq n-\deg_{alg}(f) $$ for every $y\in S$. Then
 $$\sum\limits_{y\in S}(\wt(y)+\wt(\pi(y)))\geq \sum\limits_{y\in S}(n-\deg_{alg}(f)). $$
 It is easy to see that $\sum\limits_{y\in S}(\wt(y)+\wt(\pi(y)))=2\sum\limits_{x\in
\mathbb{F}_2^n}f(x)\wt(x)$ and $\sum\limits_{y\in S}(n-\deg_{alg}(f))=\wt(f)(n-\deg_{alg}(f))$.  $\square$

We can treat Theorem \ref{ThAI3} as an estimation of mean weight of 1s of Boolean functions with known degree. 
Below we clarify details of the distribution of 1s.

Consider a vector space $\mathbb{F}_2^S$ consist of functions $h|_S$, where $h$ are $n$-variables Boolean functions. Let $\langle u,v\rangle$ be a bilinear form on   $\mathbb{F}_2^S$ defined by the equation
$$\langle u,v\rangle=\bigoplus_{x\in S}u(x)v(x).$$ This form is nondegenerate because for every $u\neq {\bf 0}$ there exists $v$ such that $\langle u,v\rangle=1$. Indeed, if $u(x_0)=1$ then we can take an indicator function of $x_0$  as $v$. Then the equation $\dim(L)+\dim(L^\bot)=|S|$ is true for every linear subspace $L\subset \mathbb{F}_2^S$.

\begin{proposition}
 Let $f:\mathbb{F}_2^n\rightarrow \mathbb{F}_2$ be a Boolean function of degree $d$ and $0\leq t \leq \min(n-d,d-1)$. Then $\sum\limits_{x, \wt(x)\leq \frac{n-d-t}{2}}f(x)\leq \sum\limits_{x, \wt(x)\geq \frac{n-d+t}{2}}f(x)$. 
\end{proposition}
Proof. 
Let $S=\supp(f)$, and let $L_m=\{h|_S : \deg_{alg}(h)\leq m\}$ denote the space of restrictions to $S$ of Boolean functions of algebraic degree at most $m$.
By Lemma \ref{lemIA2}, if $m+k<n-d$, then
$L_m\subseteq L_k^\perp$.
Since the bilinear form on $\mathbb F_2^S$ is nondegenerate,
$\dim(L_k)+\dim(L_m)\leq |S|$.

We next show that
$\{x^u|_S:u\in S,\ {\wt}(u)\leq k\}$
is a linearly independent set. Enumerate the elements of $S$ as
$S=\{u^1,\ldots,u^{|S|}\}$
in nonincreasing order of Hamming weight, with arbitrary ordering among elements of equal weight. Suppose that $\{x^{u^1}|_S,\ldots,x^{u^m}|_S\}$
is linearly independent, and put $v=u^{m+1}$. For every $j\leq m$, either
${\wt}(u^j)>{\wt}(v)$, in which case $u^j\nleq v$, or
${\wt}(u^j)={\wt}(v)$ and $u^j\neq v$, in which case again $u^j\nleq v$. Hence
$x^{u^j}(v)=0$ if  $j\leq m$,
whereas
$x^v(v)=1$.
Thus the enlarged set is linearly independent. Consequently,
$$
\dim(L_k)\geq
\bigl|\{u\in S: {\wt}(u)\leq k\}\bigr|.
$$

Therefore, whenever $m+k<n-d$,

$$
\bigl|\{u\in S: {\wt}(u)\leq k\}\bigr|+\bigl|\{u\in S: {\wt}(u)\leq m\}\bigr|
\leq |S|.
$$

Equivalently,

$$
\bigl|\{u\in S: {\wt}(u)\leq k\}\bigr|\leq \bigl|\{u\in S: {\wt}(u)\geq m+1\}\bigr|.
$$
Now put $
k=\left\lfloor\frac{n-d-t}{2}\right\rfloor$ and 
$m=n-d-k-1$.
Then $m+k=n-d-1<n-d$, and

$$
m+1=n-d-\left\lfloor\frac{n-d-t}{2}\right\rfloor= \left\lceil\frac{n-d+t}{2}\right\rceil.
$$

Since Hamming weights are integers,
${\wt}(u)\geq m+1$ if and only if ${\wt}(u)\geq\frac{n-d+t}{2}$.
The desired inequality follows. $\square$

\section{Equivalent Forms of the Main Inequality}

\begin{proposition}\label{feie11}  The inequality (\ref{feie1}) is equivalent to inequality 
$$2\sum\limits_{x\in
\mathbb{F}_2^n}f(x)\wt(x)\leq \wt(f)(n+\deg_{alg}(f)).$$
\end{proposition}
Proof.
If we change variables $x_i$ to $x_i\oplus 1$ for all $i\in 1,\dots,n$  we obtain this  inequality from (\ref{feie1})  and preserve the  set of all $n$-variable functions of degree $d$.  $\square$

We will use the following well-known property.
\begin{proposition}\label{feip5}  For any Boolean function $f: \mathbb{F}_2^n
\rightarrow \mathbb{F}_2$  and any nondegenerate affine
transformation $L:\mathbb{F}_2^n \rightarrow \mathbb{F}_2^n$ we have $\deg_{alg}(f\circ L)=\deg_{alg}(f)$.
\end{proposition}

\begin{proposition}  The inequality (\ref{feie1}) is equivalent  to inequality
\begin{equation}\label{feie2}
-\deg_{alg}(f)\wt(f)\leq2^{n-1}\sum\limits_{i=1}^nW_f(u_i)\leq \deg_{alg}(f)\wt(f),
\end{equation}
where  $\{u_1, u_2,\dots, u_n\}$ is an arbitrary  basis of $\mathbb{F}^n_2$
\end{proposition}

Proof.
It is easy to see that $\sum\limits_{i=1}^n(-1)^{x_i}=n-2\wt(x)$. Then 
$$-\deg_{alg}(f)\wt(f)\leq- \sum\limits_{x\in
\mathbb{F}_2^n}f(x)\sum\limits_{i=1}^{n}(-1)^{x_i}\leq \deg_{alg}(f)\wt(f).$$
Since $f(x)=\frac{1}{2}(1-(-1)^{f(x)})$ we have $$-\sum\limits_{x\in
\mathbb{F}_2^n}f(x)(-1)^{x_i}=
-\frac{1}{2} \sum\limits_{x\in
\mathbb{F}_2^n}(1-(-1)^{f(x)})(-1)^{x_i}=\frac{1}{2} \sum\limits_{x\in
\mathbb{F}_2^n}(-1)^{f(x)\oplus(x,e_i)}=2^{n-1}W_f(e_i),$$ where $e_i=(0,\dots,0,1,0,\dots,0)$ is a standard unit vector. Therefore, we prove  (\ref{feie2}) for the standard basis.
 By Proposition \ref{feip5} we obtain inequalities  (\ref{feie2}) are  true for every basis. By a similar way we can conclude that the  inequalities  (\ref{feie1})  follow from (\ref{feie2}). $\square$

Note that we can take $|W_f(u_i)|$ instead of  $W_f(u_i)$ in the right inequality (\ref{feie2})  because 
$W_g(z)=(-1)^{(a,z)}W_f(z)$, where $g(x)=f(x)\oplus (a,x)$.

\section{$p$-Biased Distribution}

Let $X_1,\dots, X_n$ be independent binary random variables such that $X_i=1$ with probability $p$ for $i=1,\dots,n$, where $0\leq p\leq 1$.
Define symmenric probabilistic distribution on $\mathbb{F}_2^n$ by the formula: 
${\rm Pr}\{X=x\}=p^{\wt(x)}(1-p)^{n-\wt(x)}$. For function $f:\mathbb{F}_2^n\rightarrow \mathbb{F}_2$ define 
 $$\mu_f(p)={\rm Pr}(\supp(f))={\mathsf E}(f)=\sum\limits_{x\in
\mathbb{F}_2^n}f(x){\rm Pr}\{X=x\}.$$
 It is easy to see that $\mu_{f\oplus 1}(p)=\mu_f(1-p)$. So, we will consider only $0\leq p\leq 1/2$.
By definition, we have $\mu_f(1/2)=\frac{\wt(f)}{2^n}$. It is well-known that 
\begin{equation}\label{fei4}
\wt(f)\geq 2^{n-\deg_{alg}(f)}
\end{equation}
 if $f$ is not a zero constant.  This bound is known as a minimal code distance of Reed--Muller code.  In other words this bound implies  $\mu_f(1/2)\geq\frac{1}{{2^{\deg_{alg}(f)}}}$ (Schwartz--Zippel lemma).
Recently this inequality was generalized by the following way. 

\begin{proposition}[\cite{Dinur}, Claim 6.8]\label{Dimur}
For $0\leq p\leq 1/2$ and every Boolean function $f$, $f\neq {\bf 0}$, it holds 
 $\mu_f(p)\geq p^{\deg_{alg}(f)}$.
\end{proposition}

We propose further improvement of this inequality. 

\begin{proposition}\label{pAI4} Let  $0\leq p,q\leq 1/2$ and let  $f$, $f\neq {\bf 0}$, be a Boolean function $f$ of degree $d$. It holds 
 $\mu_f(pq)\geq q^d\mu_f(p)$.
 \end{proposition}
 Proof. 
 Let $X_1,\dots, X_n, Y_1,\dots, Y_n$ be independent binary  random variables such that $X_i=1$ with probability $p$ and where $Y_i=1$ with probability $q$, where $0\leq p,q\leq 1/2$. Then $X_iY_i$ are independent binary  random variables with ${\rm Pr}\{X_iY_i=1\}$ for $i=1,\dots,n$.
 
Let $x\cdot y= (x_1y_1,\dots, x_ny_n)$. We have
 $$\mu_f(pq)=\sum\limits_{x,y\in\mathbb{F}_2^n}f(xy){\rm Pr}\{XY=xy\}=\sum\limits_{x,y\in\mathbb{F}_2^n}f(xy){\rm Pr}\{XY=xy| X=x\}{\rm Pr}\{X=x\}$$ $$=\sum\limits_{x\in\mathbb{F}_2^n}{\rm Pr}\{X=x\}\sum\limits_{y\in\mathbb{F}_2^n}
f(xy){\rm Pr}\{XY=xy| X=x\}.$$
Define functions $f_x(y)$ by the equation $f_x(y)=f(xy)$. By definition $\deg_{alg}(f_x)\leq \deg_{alg}(f)$.  Moreover, $f_x\neq {\bf 0}$ if $f(x)=1$. By Proposition \ref{Dimur} we obtain that $\mu_{f_x}(q)\geq q^d$.
If $f(x)=1$ then
$$\sum\limits_{y\in\mathbb{F}_2^n}
f(xy){\rm Pr}\{XY=xy| X=x\}=\mu_{f_x}(q)\geq q^d.$$
Then 
 $$\mu_f(pq)=\sum\limits_{x\in\mathbb{F}_2^n}{\rm Pr}\{X=x\}\sum\limits_{y\in\mathbb{F}_2^n}
f(xy){\rm Pr}\{XY=xy| X=x\}\geq q^d\sum\limits_{x\in\mathbb{F}_2^n}f(x){\rm Pr}\{X=x\}= q^d\mu_f(p).$$
 $\square$

\begin{corollary}\label{corAI1}
For $0\leq s\leq 1/4$ and every $n$-variable Boolean function $f$, $f\neq {\bf 0}$, of degree $d$, it holds 
 $\mu_f(s)\geq\frac{s^{d}\wt(f)}{2^{n-d}}$.
 \end{corollary}
Proof. We use Proposition \ref{pAI4} with $p=1/2$ and $q=2s$.
 $\square$

\begin{proposition}\label{pAI5}
For $0\leq p\leq 1/2$ and every Boolean function $f$, $f\neq {\bf 0}$, of degree $d$ it holds\\
 $\mu_f(p)\geq \wt(f)p^{(n+d)/2}(1-p)^{(n-d)/2}$.
\end{proposition}

Proof. The function $h(t)=\alpha^t$ is convex. Thus, by Jensen's inequality we obtain that 
$$\frac{1}{\wt(f)}\sum\limits_{x\in \supp(f)}\left(\frac{p}{1-p}\right)^{\wt(x)}\geq \left(\frac{p}{1-p}\right)^{\sum\limits_{x\in \supp(f)}\frac{\wt(x)}{\wt(f)}}.$$
Since $\frac{p}{1-p}\leq 1$ by Proposition \ref{feie11} we have $$ \left(\frac{p}{1-p}\right)^{\sum\limits_{x\in \supp(f)}\frac{\wt(x)}{\wt(f)}}\geq  \left(\frac{p}{1-p}\right)^{\frac{n+d}{2}}.$$
Then $$\mu_f(p)=\sum\limits_{x\in \supp(f)}p^{\wt(x)}(1-p)^{n-\wt(x)}\geq wt(f)p^{(n+d)/2}(1-p)^{n-(n+d)/2}.$$  $\square$

Note that  
\begin{equation}\label{efei5}
\frac{p^d}{2^{n-d}}\geq p^{(n+d)/2}(1-p)^{(n-d)/2}
\end{equation}
 and, consequently, the bound  from Corollary \ref{corAI1} is better than the bound from Proposition \ref{pAI5}. Indeed, (\ref{efei5}) is equivalent to $1\geq 4p(1-p)$.  However, the bound  from Corollary \ref{corAI1} is valid only if $p\leq 1/4$.  For the case $1/4\leq p\leq 1/2$, the bound from Proposition \ref{pAI5} is better than the bound from Proposition \ref{Dimur} if
 $p^d< \wt(f)p^{(n+d)/2}(1-p)^{(n-d)/2}$, i. e., $\wt(f)\geq (p(1-p))^{(d-n)/2}$. For $p=1/2$ the last equation coincide with (\ref{fei4}), so itis holds for all functions of degree $d$.

\section{Acknowledgements}

The author thanks Storozhuk for his continued interest in his work. The  main inequality was proved  with the assistance of ChatGPT.


\begin{thebibliography}{9}

\bibitem {Alon}
 Alon N., Kaufman T.,  Krivelevich M., Litsyn S., and  Ron D., Testing Reed-Muller codes, IEEE Trans. Inform. Theory, vol. 51 (2005), pp. 4032–4039. https://doi.org/10.1109/TIT.2005.856958, 


\bibitem {Carlet}
 Carlet C.,  \emph{Boolean Functions for Cryptography and Coding Theory}.
  Cambridge University Press, 562 pages, 2020.

\bibitem {Dinur}
 Dinur I.,  Filmus Y.,  Harsha P., Agreement tests on graphs and hypergraphs,
SIAM Journal on Computing, vol. 54 iss. 2 (2025). https://doi.org/10.1137/21M1397684



\bibitem {FG}
 Friedgut E.,  Kalai G., Every monotone graph property has a
sharp threshold. Proceedings of the American mathematical Society,
124(10) (1996), pp. 2993-3002.

\bibitem {Stanica}
Gangopadhyay S.,  Stănică P.  Fourier entropy-influence conjecture for cryptographic Boolean functions. Trans. Adv. Res.(Special issue on Advances in Cryptology and Information Security), vol. 12(2), (2016), pp. 8-14. 

\bibitem {Li-Tang}
 Li Z.,  Tang D.,
A note on the FMEI of the Boolean functions in the Generalized Maiorana-McFarland construction,
Discrete Applied Mathematics,
vol. 387 (2026), pp. 106-115.
https://doi.org/10.1016/j.dam.2026.02.052.


\bibitem {Pot}
 Potapov V.N., On the number of relevant variables for discrete functions, Cryptography and Communications, vol.17 (2025),  pp. 989–998.  https://doi.org/10.1007/s12095-025-00797-4

\bibitem {Sala}
Sălăgean A., Reyes-Paredes P., Bounds for the average degree-$k$ monomial density of Boolean functions. 
Cryptogr. Commun.  vol. 18, no. 1  (2026), pp. 187-211. https://doi.org/10.1007/s12095-025-00839-x








\end{thebibliography}
\end{document}